\documentclass[12pt,a4paper]{article}  
\usepackage{a4wide}
\usepackage{graphicx}
\usepackage{subfigure}
\usepackage[centertags]{amsmath}
\usepackage{amssymb}
\usepackage{cite}
\usepackage{epsfig}
\usepackage{longtable}
\usepackage{enumerate}
\usepackage{scrtime}

\newcommand{\id}{1\!\!1}

\begin{document}  

\thispagestyle{empty}

\begin{flushright}  
                     IPPP/05/09\\
                     DCPT/05/18\\
\end{flushright}  
\vskip 2cm    

\begin{center} 
{\huge Geometry of Rank Reduction}   
\vspace*{5mm} \vspace*{1cm}   
\end{center}  
\vspace*{5mm} \noindent  
\vskip 0.5cm  
\centerline{\bf Stefan F\"orste$^a$, Hans Peter Nilles$^{b,c}$, Ak\i{}n
  Wingerter$^b$}
\vskip 1cm
\centerline{$^a$
\em Institute for Particle Physics Phenomenology (IPPP)}
\centerline{\em South Road, Durham DH1 3LE, United Kingdom}
\vskip 0.5cm
\centerline{$^b$\em Physikalisches Institut, Universit\"at Bonn}
\centerline{\em Nussallee 12, D-53115 Bonn, Germany}
\vskip 0.5cm
\centerline{$^c$ \em Theory Division, Physics Department}
\centerline{\em CERN, CH-1211 Geneva 23, Switzerland}
\vskip2cm

\centerline{\bf Abstract}  
\vskip .3cm

We introduce continuous Wilson lines to reduce the rank of the gauge group in orbifold
constructions. In situations where the orbifold twist can be realised
as a rotation in the root lattice of a grand unified group we derive
an appealing geometric picture of the symmetry breakdown. This symmetry
breakdown is smooth and corresponds to a standard field theory Higgs mechanism.
The embedding into heterotic string theory is discussed.

\vskip .3cm

\newpage  

\section{Introduction}

To establish a connection between superstring theory in $d=10$ space-time dimensions
and particle physics models in $d=4$ we need to have a detailed understanding of
the compactification of extra space dimensions. Over the years, orbifold compactification 
\cite{Dixon:1985jw,Dixon:1986jc} turned out to be a useful tool for string
model building. This scheme combines the technical simplicity of torus compactification
with the requirements of realistic gauge group and particle spectrum. With the inclusion
of background fields \cite{Ibanez:1986tp,Ibanez:1987pj} (as e.g.\ Wilson lines), a large
number of models can be constructed, most notably in the framework of heterotic string
theory. Although simplified, the scheme leaves open so many possibilities that, 
at present, a full
classification seems to be hopeless.

More recently, the question of gauge unification in higher dimensions has been studied
in $d=5$ \cite{Kawamura:2000ev} and $d=6$ \cite{Asaka:2001eh} in a pure field theoretical 
framework (so-called orbifold GUTs). Some of the successful aspects of string orbifold
models, as e.g.\ the doublet-triplet splitting \cite{Ibanez:1987sn}
in grand unified theories, can be incorporated 
in this scheme as well. The process of compactification is further simplified and allows
more flexibility in model building, since the severe consistency conditions of
string theory are not taken into account. The field theory orbifold GUTs should therefore
be understood as a set-up for model building where questions of consistency of the
higher dimensional quantum field theory have been postponed. Such a consistency
might be achieved a posteriori by a suitable ultraviolet completion, 
ultimately through an 
embedding in a consistent string theory along the lines of
\cite{Kobayashi:2004ud,Forste:2004ie,Kobayashi:2004ya,Buchmuller:2004hv,Faraggi:2005ib}.

The mechanism of orbifold compactification (whether in $d=10$, 5, or whatever) might,
of course, lead to results that could be considered as an artifact of its
simplicity. We think that the question about the rank of the gauge group might fall
in this category. In contemporary constructions the orbifolding procedure does not
lower the rank of the higher dimensional grand unified gauge group. The present paper
is devoted to the study of rank reduction in orbifold compactifications. Earlier 
work in that direction can be found in \cite{Ibanez:1987xa} in the framework of the
$\mathbb{Z}_3$ orbifold, where, unfortunately, it was difficult to make contact to
even semi-realistic models of particle physics in the rank reduced case.

This mechanism of rank reduction requires a more sophisticated treatment of the
orbifolding procedure than usually employed. The space-time twist has to be presented
as a rotation (and not just as a shift) in the root lattice of the higher
dimensional gauge group. Certain continuous Wilson lines can then lead to a
reduction of the rank of that group. From a low-energy effective field theory
point of view such a Wilson line would correspond to a non-vanishing vacuum
expectation value of an untwisted (bulk) scalar field
\cite{Font:1988tp,Font:1988mm} and represents thus a stringy implementation of
the Higgs mechanism.

As the description of the mechanism is technically quite complicated, we shall
not give the discussion in full generality but shall instead concentrate
on a specific (though not simple) example: $\text{E}_6$ gauge symmetry in two 
extra dimensions. A more complete treatment can be found in
\cite{WingerterPhD:2005}.

In section 2 we shall give a short introduction to the orbifold
technology: twists, shifts, discrete and continuous Wilson lines as well
as the basic picture of rank reduction. Section 3 will contain a detailed
discussion of Weyl rotations in $\text{E}_6$, an example with a breakdown
to $\text{SO}(10)\times \text{U}(1)$ and the action of continuous Wilson lines. In
section 4 we provide an explicit example of a $\mathbb{Z}_2$ orbifold in
$d=6$ with gauge group $\text{E}_6$ in the bulk. With a discrete Wilson line 
we can break to the Pati-Salam Group, while a (rank reducing)
continuous Wilson line breaks the gauge group
$\text{SU}(4)\times \text{SU}(2)\times \text{SU}(2)$ to the standard model
$\text{SU}(3)\times \text{SU}(2)\times \text{U}(1)$. Section 5 discusses the possible embedding
of the $d=6$ model in the full $d=10$ string setup. We shall see that
the model of section 4 might find its completion in one of the  models  discussed in
ref. \cite{Kobayashi:2004ya} or a variant thereof. We then analyse
the lessons one might learn from such an embedding. Section 6
gives an outlook and conclusions.

\section{Orbifold Constructions in Six Dimensions}

We will briefly review orbifold constructions \cite{Dixon:1985jw,
  Dixon:1986jc}.  
Starting with six dimensions, two dimensions are
compactified on an orbifold 
\begin{equation}
\mathcal{O} = T^2 \big{/} P.
\end{equation}
An orbifold is defined to be the quotient of a torus over a
discrete set of isometries, called the {\it point group}
$P$. Alternatively, one can start
with the complex plane, and first 
identify points which differ by translations (lattice shifts) $L$ in
order to arrive at the torus, 
and then mod out the action of $P$: 
\begin{equation}
T^2 = \mathbb{C} \big{/} L \quad \leadsto \quad \mathcal{O} =
T^2 \big{/} P = \mathbb{C} \big{/} S. 
\end{equation}
$S$ is called the {\it space group}, and is the semidirect product of
the point group $P$ and the translation group $L$ defining the
torus. For the action of the point group to be well-defined, elements
of $P$ must be automorphisms of the lattice defining the torus. 

The original theory in six dimensions is taken to be a grand unified gauge theory.
The action of the point group on the space-time degrees of freedom is
generically accompanied by an action on the gauge degrees of
freedom, $P \hookrightarrow G$, where the embedding is a
homomorphism. $G$ is a subgroup of the automorphisms of the Lie
algebra $\mathfrak{g}$ describing the gauge symmetry, and is called
the {\it gauge twisting group}.

\subsection{Embedding the Twist in the Gauge Degrees of Freedom}

Any inner Lie algebra automorphism $\sigma$ of finite order $N$ can be
realised as a shift $X \mapsto X + V$, in the root lattice $\Lambda$ of $\mathfrak{g}$ such
that its action on the step operators corresponding to the simple
roots $\alpha_k$ and the extended root $\alpha_0$ is given by
\cite{Kac:1969xxx1} 
\begin{equation}
\sigma\left( E_{\alpha_k}\right) = \exp\left( 2\pi
i \alpha_k \cdot V\right) E_{\alpha_k}, \quad k=0,\ldots, \text{rank
}\mathfrak{g}, 
\end{equation}
with $\alpha_0 =-\sum_{j=1}^{ \text{rank
}\mathfrak{g}} \alpha_j k_j $, where $k_j$ are the Kac labels. On
the Cartan generators $H_i$, 
the action of $\sigma$ is trivial. Thus none of the
Cartan generators is projected out. It is therefore clear
that by this construction the rank of the algebra cannot be reduced.

To be able to reduce the rank of the gauge group we need
an alternative approach to symmetry breaking in which the
action of the twist in the gauge algebra transforms some of
the Cartan generators non-trivially.  
Such a mechanism can be realised as follows. As the elements of $P$
are automorphisms of the lattice defining $T^2$, it is
natural to associate with them automorphisms of the root lattice
$\Lambda$, i.e.~to realise the twist in the space-time as a twist in
the gauge degrees of freedom \cite{Ibanez:1987xa}. The automorphisms
of the root lattice, the Weyl group $\mathcal{W}$ of $\mathfrak{g}$,
is generated by the reflections  
\begin{equation} 
r_{\alpha_k} : \xi \mapsto \xi - 2 \frac{\langle\alpha_k, \xi \rangle}{\langle\alpha_k, \alpha_k \rangle} \alpha_k, 
\end{equation}  
where the $\alpha_k$ are the simple roots of $\mathfrak{g}$. There is
a natural lift of $r_{\alpha_k}$ to the Lie algebra $\mathfrak{g}$
given by \cite{Humphreys:1980dw, Freundethal:1969hn, Schellekens:1987ij} 
\begin{equation}
\tilde{r}_{\alpha_k} = \exp\left( \frac{i\pi}{2}\left(E_{\alpha_k} +
E_{-\alpha_k} \right) \right), 
\label{eq:shiftless_lift}
\end{equation}
so that the lift $\tilde{w}$ of an arbitrary element of $w \in
\mathcal{W}$ is given by the product of lifts of simple Weyl
reflections. Under the lift of a single Weyl reflection $r_\alpha$,
the generators of the Lie algebra transform as 
\begin{equation}
\renewcommand{\arraystretch}{1.5}
\begin{array}{llll}
\tilde{r}_\alpha \, (\lambda \cdot H) \, \tilde{r}_\alpha^{-1} =
(r_\alpha(\lambda))\cdot H,\\ 
\tilde{r}_\alpha E_\beta \tilde{r}_\alpha^{-1} = c_\alpha(\beta)\,
E_{r_\alpha(\beta)}.
\label{eq:neat_form_for_transformation_properties_of_operators}
\end{array}
\end{equation}
For an explicit calculation, the complex phases $c_\alpha(\beta)$ must be
determined. In appendix \ref{sec:calculation_of_phase}, we 
express these phases in terms of the structure constants of the algebra.

As the order of the Weyl group is finite, and the automorphism group of a Lie algebra is itself a Lie algebra, it is clear that the relation between the lift described above and the automorphisms realised by shifts cannot be one-to-one. Refs.~\cite{Hollowood:1987hf, Schellekens:1987ij, Bouwknegt:1988hn} are concerned with determining the shift vectors corresponding to the conjugacy classes of the Weyl group.

The definition of the lift as given by eq.~(\ref{eq:shiftless_lift}) is not unique. 
One can also first shift the lattice by an arbitrary vector $v$ before 
twisting it with $\tilde{r}_{\alpha_k}$ \cite{Freundethal:1969hn, Schellekens:1987ij} : 
\begin{equation}
\tilde{r}_{\alpha_k}' = \tilde{r}_{\alpha_k} \exp\left( 2\pi v\cdot H \right) = \exp\left( \frac{i\pi}{2}\left(E_{\alpha_k} + E_{-\alpha_k} \right) \right)  \exp\left( 2\pi v\cdot H \right).
\label{eq:lift_with_lift}
\end{equation}
In the following, we shall see that this generalisation will lead to many new
possibilities for model building.

A nontrivial consistency condition is that the order of the
algebra automorphism should be a divisor of the order of the point
group. Note that the Cartan generators transform
non-trivially, and step operators need not be eigenstates under the
orbifold action. Starting from a set of generators in the Cartan-Weyl
basis yields a set of invariant generators which are
typically not in the Cartan-Weyl basis of the unbroken algebra. The
invariant generators consist of the sum of a generator and its
images. In order to find the Cartan-Weyl basis of the
unbroken algebra one proceeds as follows. First, we identify a Cartan
subalgebra by taking the (linear combinations of) Cartan generators of
the original algebra 
which are invariant under the orbifold action. These we supplement by
linear combinations of step operators which are even under the
orbifold group and are not charged under the invariant Cartan
generators. If the number of invariant Cartan generators differs from
the rank of the original gauge symmetry by more than one we need to
supplement by more than one linear combination of step
operators. These should mutually commute.

After a Cartan subalgebra is chosen one has to simultaneously
diagonalise its adjoint action on the remaining invariant combinations
of step operators. Eventually, this procedure leads to the unbroken
gauge symmetry written in the Cartan-Weyl basis allowing for an
identification of the group from the Cartan matrix or the Dynkin
diagram. 

Once we have clarified this, we shall see that the rank of the
gauge group still remains the same. Some of the Cartan operators
in the original gauge group have been projected out, but some new
ones (linear combinations of the step operators) appear and will
replace the former ones. This is a result of the fact that, if one
just considers the point group and not the full space group of the
orbifold, any rotation in the gauge group can be represented by
a shift \cite{Ibanez:1987pj}. 
Rank reduction needs more than just this. It also needs
a representation of the full space group in the gauge group,
thus additional Wilson lines.

\subsection{Wilson Lines}

So far
we have embedded the twist in  space-time as a twist in the gauge degrees of freedom. 
Analogously, each shift in the space-time defining the torus can be associated with a 
shift in the co-root lattice\footnote{For algebras of type $ADE$, the
  co-root lattice is equal to the root lattice.} of $\mathfrak{g}$, and
this corresponds to a Wilson line $W$.  
Around non-contractible loops, the operators will then transform with a phase,
\begin{equation}
E_\alpha \rightarrow \exp\left(2\pi i \alpha \cdot W\right) E_\alpha.
\label{eq:trafo_under_wilsonline}
\end{equation}
A Wilson line might thus remove some of the step operators $E_\alpha$. If it projects out step
operators that play the role of Cartan operators in the ``twisted'' gauge group the rank of
the gauge group can thus be reduced.

When considering Wilson lines in the presence of the space-time twist realised as a rotation, there are 3 cases to be distinguished:
\begin{enumerate}[(i)]
\item $W$ is left invariant by the twist $s$,
\item $W$ is completely rotated by the twist $s$,
\item Some of the components of $W$ are rotated by the twist $s$.
\end{enumerate}
In the following, we will concentrate on the first 2 cases. Consider the case, when the Wilson line is invariant, and for the sake of briefness, assume that the twist is of order 2. Applying twice the same gauge transformation must act as the identity. Denoting the gauge degrees of freedom by $X$, we have
\begin{equation}
X \enspace \overset{s}{\rightarrow} \enspace  s(X) + W \enspace \overset{s}{\rightarrow} \enspace s^2(X) + s(W) + W = X + 2W,
\end{equation}
where we used the fact that $s$ is of order 2, and leaves $W$
invariant. We conclude that $2W$ must be in the co-root lattice, and
hence is discrete. 

Let us now consider the case, when the Wilson line is completely rotated by the twist $s$, and repeat the previous arguments:
\begin{equation}
X \enspace \overset{s}{\rightarrow} \enspace  s(X) + W \enspace
\overset{s}{\rightarrow} \enspace s^2(X) + s(W) + W = X - W + W = X .
\end{equation}
In contrast to the case, where the Wilson line was invariant, $W$ is now not restricted to lie in a lattice, and hence can be rescaled by an arbitrary real parameter $\lambda$.

The step operators are still subject to the transformation law given in eq.~(\ref{eq:trafo_under_wilsonline}) around non-contractible loops, but this time the continuous parameter $\lambda$ in
\begin{equation}
E_\alpha \rightarrow \exp\left(2\pi i \alpha \cdot \lambda \, W\right) E_\alpha
\label{eq:trafo_under_cont_wilsonline}
\end{equation}
will have the effect that there will always be a non-trivial phase, if $\alpha \cdot W \neq 0$. 
Hence, the corresponding operators will be projected out and this leads to rank reduction.
In section \ref{sec:discuss_reducing_rank} we shall discuss this in detail in an explicit
example.

So far our technical treatment of the Wilson lines. As this is the central point of the
mechanism of rank reduction we add a more intuitive discussion of the action of
Wilson lines in this set-up. It will be useful to make contact to the picture from
the effective low-energy field theory point of view.
Wilson lines stand for vacuum
expectation 
values of an internal component of the gauge field. The 
two qualitatively different types of Wilson lines will be discussed in
the following.

\subsubsection{Discrete Wilson Lines: No Rank Reduction}

First we consider the case that the vev of an internal gauge field
component points into a Lie algebra direction which is even under the
orbifold group. Without loss of generality the vev can be taken to
point into invariant Cartan directions ($m$ labels a compactified
space direction and $I$ a Cartan direction)
\begin{equation}
A_m = A_m^I H^I .
\end{equation}
We associate this vev with a vector $W_i$ of the maximal torus according
to \cite{Ibanez:1986tp}.
\begin{equation}\label{wilsonline}
\oint_i A_m^I dz^m = 2\pi A_m ^I e_i^m = 2\pi W_i ^I,
\end{equation}
where $i$ labels a non contractible loop on $T^2$, i.e.\ $e_i$ is a
basis vector in the $T^2$ lattice. The embedding of the point group
into the gauge group was realised as an automorphism of the root
lattice which is the lattice defining the maximal torus (for simply
laced groups). Hence it
induces a nontrivial transformation of the vector $W_i$. The
condition that the vev points into an invariant direction means that
the vector $W_i$ has to be chosen such that it is invariant under the
orbifold action.

Due to the internal vector index the vev is odd under
the full action of the orbifold and the corresponding moduli are
projected out. The Wilson line and its orbifold image should be
identified by symmetries of the theory. The Wilson line is
quantised or {\it discrete}. In our case the scalar product of $2W_i$
with all occurring 
weight vectors has to be integer. From the way we constructed the
Cartan subalgebra of the unbroken gauge group in the previous section
it is clear that the discrete Wilson line commutes with all elements
of the Cartan subalgebra. Hence the rank is not reduced. 

\subsubsection{Continuous Wilson Lines: Rank Reduction}

Here, we consider the option that the Wilson line points into a
direction which is not invariant under the orbifold action. By
employing unbroken gauge transformations the Wilson line can be
always chosen to point into odd Cartan directions. According to
eq.~(\ref{wilsonline}) it can again be associated with a vector on the
maximal torus. This vector is taken to transform non trivially under
the embedding of the point group into the gauge group. 
Because of the internal vector index $m$ the overall
transformation under the point group is even, the vev corresponds to a 
modulus, the Wilson line is {\it continuous}.

Since the continuous Wilson line points into an odd Cartan direction
it can reduce the rank. The reason can be seen by looking again at our
construction of the Cartan algebra in the unbroken gauge group. The
combinations of step operators supplementing the even Cartan
generators can be charged under the odd Cartan generators. Hence it
can happen that an element of the Cartan algebra in the unbroken gauge
group does not commute with the continuous Wilson line and, accordingly,
is projected out. The rank of the gauge group is reduced.

Since continuous Wilson lines are associated with moduli they should
correspond to flat directions in the four dimensional picture. This is
indeed the case. The internal gauge field components give rise to
scalars taking values in the complement of the unbroken gauge group. 
Giving non vanishing vacuum expectation values to these scalars is the
four dimensional picture for switching on a continuous Wilson line.
This symmetry breaking is thus smooth and  corresponds to a
Higgs-mechanism in the low energy-effective theory.

\section{The Breaking of $\boldsymbol{\text{E}_6}$}

Let us now consider  $\text{E}_6$ as an explicit example. It appears at an intermediate stage in many of
the phenomenologically interesting models derived from the heterotic $\text{E}_8\times \text{E}_8$ string theory. 
The Weyl group of $\text{E}_6$ has 51,840 elements, each of which is in one of 25 conjugacy classes \cite{Carter1972ar}. In tab.~\ref{tab:conjugacy_classes_E6}, we list for each conjugacy class one representative in terms of simple Weyl reflections \cite{mapleWeyl}, its order on the root lattice, the order of the corresponding lift to the algebra, and the associated symmetry breaking.

\begin{table}[hp]
\centering
\normalsize
\renewcommand{\arraystretch}{1.4}
\begin{tabular}{|c|p{7cm}|c|c|c|l|}
\hline
\multicolumn{1}{|c|}{No.} & \multicolumn{1}{|c|}{Weyl Group Element} & \multicolumn{1}{|c|}{ord${}_1$} & \multicolumn{1}{|c|}{ord${}_2$}  & \multicolumn{1}{|c|}{Inv.}  & \multicolumn{1}{|c|}{Gauge group} \\
\hline
\hline
1 & $\id $							& 1 &  1 & 78  & $\text{E}_6$\\  
2 & $r_1$ 							& 2 &  4 & 36  & $\text{SU}(6)\times\text{U(1)}$\\ 
3 & $r_1$\! $r_2$ 						& 3 &  3 & 36  & $\text{SU}(6)\times\text{U(1)}$\\ 
4 & $r_1$\! $r_6$ 						& 2 &  4 & 30  & $\text{SO}(8)\times\text{U}(1)^2$\\ 
5 & $r_1$\! $r_2$\! $r_3$ 					& 4 &  8 & 18  & $\text{SU}(4)\times\text{U}(1)^3$\\ 
6 & $r_6$\! $r_2$\! $r_3$\! $r_6$\! $r_4$\! $r_3$		& 4 &  4 & 20  & $\text{SU}(3)^2\times\text{SU(2)}\times\text{U}(1)$\\ 
7 & $r_1$\! $r_6$\! $r_2$					& 6 &  12 & 18  & $\text{SU}(4)\times\text{U}(1)^3$\\ 
8 & $r_6$\! $r_2$\! $r_3$\! $r_4$ 				& 6 &  6 & 18  & $\text{SU}(3)^2\times\text{U}(1)^2$\\ 
9 & $r_1$\! $r_6$\! $r_2$\! $r_3$ 				& 5 &  5 & 18  & $\text{SU}(4)\times\text{U}(1)^3$\\ 
10 & $r_1$\! $r_6$\! $r_2$\! $r_3$\! $r_6$\! $r_4$\! $r_3$ 	& 12 & 24  & 6  & $\text{U}(1)^6$\\ 
11 & $r_1$\! $r_6$\! $r_2$\! $r_1$\! $r_3$\! $r_6$\! $r_2$\! $r_3$\! $r_4$\! $r_3$\! $r_5$\! $r_4$ 	& 6 &  6 & 12  & $\text{SU}(2)^3\times\text{U}(1)^3$\\ 
12 & $r_1$\! $r_6$\! $r_4$ 					& 2 &  4 & 20  & $\text{SU}(3)^2\times\text{SU}(2)\times\text{U}(1)$\\ 
13 & $r_1$\! $r_6$\! $r_3$\! $r_4$ 				& 4 &  8 & 12  & $\text{SU}(3)\times\text{U}(1)^4$\\ 
14 & $r_1$\! $r_6$\! $r_2$\! $r_3$\! $r_4$ 			& 8 &  8 & 10  & $\text{SU}(2)^2\times\text{U}(1)^4$\\ 
15 & $r_1$\! $r_2$\! $r_4$\! $r_5$ 				& 3 &  3 & 30  & $\text{SO}(8)\times\text{U}(1)^2$\\
16 & $r_1$\! $r_2$\! $r_3$\! $r_4$\! $r_5$ 			& 6 &  12 & 8  & $\text{SU}(2)\times\text{U}(1)^5$\\ 
17 & $r_1$\! $r_6$\! $r_2$\! $r_3$\! $r_6$\! $r_4$\! $r_3$\! $r_5$ 		& 9 &  9 & 8  & $\text{SU}(2)\times\text{U}(1)^5$\\ 
18 & $r_1$\! $r_6$\! $r_2$\! $r_4$ 						& 6 &  12 & 12  & $\text{SU}(2)^3\times\text{U}(1)^3$\\ 
19 & $r_1$\! $r_6$\! $r_2$\! $r_3$\! $r_4$\! $r_5$ 				& 12 &  12 & 6  & $\text{U}(1)^6$\\ 
20 & $r_1$\! $r_6$\! $r_2$\! $r_3$\! $r_5$					& 10 &  20 & 8  & $\text{SU}(2)\times\text{U}(1)^5$\\ 
21 & $r_6$\! $r_2$\! $r_3$\! $r_6$\! $r_2$\! $r_3$\! $r_4$\! $r_3$\! $r_6$\! $r_2$\! $r_3$\! $r_4$ 	& 2 &  2 & 38  & $\text{SU}(6)\times\text{SU(2)}$\\ 
22 & $r_1$\! $r_6$\! $r_2$\! $r_3$\! $r_6$\! $r_2$\! $r_3$\! $r_4$\! $r_3$\! $r_6$\! $r_2$\! $r_3$\! $r_4$ 	& 4 &  8 & 14  & $\text{SU}(2)^4\times\text{U}(1)^2$\\ 
23 & $r_1$\! $r_6$\! $r_2$\! $r_4$\! $r_5$ 					& 6 &  12 & 12  & $\text{SU}(2)^3\times\text{U}(1)^3$\\ 
24 & $r_1$\! $r_6$\! $r_2$\! $r_3$\! $r_6$\! $r_2$\! $r_3$\! $r_4$\! $r_3$\! $r_6$\! $r_2$\! $r_3$\! $r_4$\! $r_5$ 	& 6 &  6 & 14  & $\text{SU}(2)^4\times\text{U}(1)^2$\\ 
25 & $r_1$\! $r_6$\! $r_2$\! $r_1$\! $r_3$\! $r_6$\! $r_2$\! $r_1$\! $r_3$\! $r_4$\! $r_3$\! $r_6$\! $r_2$\! $r_1$\! $r_3$\! $r_4$\! $r_5$\! $r_4$\! $r_3$\! $r_6$\! $r_2$\! $r_3$\! $r_4$\! $r_5$	& 3 &  3 & 24  & $\text{SU}(3)^3$\\ 
\hline
\end{tabular}
\caption{Weyl group conjugacy classes of $\text{E}_6$ and the
  associated symmetry breakings. The entry of `ord${}_1$' gives the order of
  the automorphism on the root lattice whereas `ord${}_2$' gives the order
  of the induced algebra automorphism when the shiftless lift of equation
  (\ref{eq:shiftless_lift}) is applied. In column `Inv.' the number of
  invariant elements is reported.} 
\label{tab:conjugacy_classes_E6}
\end{table}

Naively one might think that with this, all the  
possibilities of breakdown of $\text{E}_6$ would be classified.
But this is actually not the case. As we have explained in the last section, we can 
generalise the lift (cf.~eq.~(\ref{eq:lift_with_lift})) by first shifting the 
lattice by an arbitrary vector $v$ before twisting it 
by an element of one of the conjugacy classes.
This opens up many more possibilities which,
unfortunately, are difficult to classify in full generality.

In fact we shall here consider an example with a generalised lift.
Following the argumentation of ref. \cite{Nilles:2004ej,Forste:2004ie} 
we are particularly interested in an $\text{SO}(10)$ gauge symmetry at
some intermediate stage, and $\text{SO}(10)$ is not 
included in the list of possible symmetry breakings of table 1. 
Introducing an appropriate shift, the fourth conjugacy class in tab.~\ref{tab:conjugacy_classes_E6} 
will correspond to a breakdown of the gauge symmetry $\text{E}_6 \rightarrow \text{SO}(10)\times \text{U}(1)$, 
and moreover, the order of the twist in the root lattice and of its lift will coincide, 
which is required for the consistency of the construction.

\subsection{Breaking $\boldsymbol{\text{E}_6}$ to $\boldsymbol{\text{SO}(10)\times\text{U}(1)}$}
\label{sec:breaking_E6_to_SO10}

For our purposes, $\text{E}_6$ is best described in terms of its embedding in $\text{E}_8$, cf.~appendix \ref{sec:choice_roots}. We consider the fourth conjugacy class $s \equiv r_1 r_6$ which is of order 2. The lift of $s$ will be given by
\begin{equation} 
\tilde{s} \equiv \tilde{r}_1 \tilde{r}_6 \exp\left( 2\pi v\cdot H\right), \quad v = \left( 0,\, 1/4,\, 0,\, 0,\, 0,\, 1/4,\, 0,\, 0 \right),
\label{eq:shivt}
\end{equation}
where the choice of $v$ is motivated by the considerations discussed
above. Using
eq.~(\ref{eq:neat_form_for_transformation_properties_of_operators}),
it is straightforward to determine the images of the 6 Cartan and 72
step operators of $\text{E}_6$ under the transformation. From the
step operators 12
are invariant, and the rest pair up to give 30 invariant
combinations. Of the 6 Cartan generators, 2 are invariant, and 4 pair
up to give 2 invariant combinations. Thus, 2 of the original Cartan
generators are projected out. Looking for a maximal commuting
subalgebra, we find that there are 2 invariant combinations of step
operators, namely $E_2+E_3$, and $E_{37}+E_{40}$, which commute with
the 4 original Cartan generators, forming the Cartan subalgebra of the
unbroken gauge group. The 46 invariant combinations are summarised in
tab.~\ref{tab:invariant_combinations_twist} in appendix
\ref{sec:calculational_breaking_E6_to_SO10}. 
      
Knowing the dimension and rank of the unbroken gauge group, it is not difficult to conclude that it corresponds to the subgroup $\text{SO}(10)\times \text{U}(1)$ of $\text{E}_6$. It will prove useful to verify this conclusion by an explicit calculation, using Dynkin's approach to group theory. Even though the dimension and the rank may uniquely specify the gauge group in this particular case, as soon as the dimension becomes small, ambiguities arise, necessitating a more thorough investigation. We present the details of the calculation in appendix \ref{sec:calculational_breaking_E6_to_SO10}.

The 32 linear combinations of operators, which are not invariant under the gauge twist, but transform with a minus sign (cf.~tab.~\ref{tab:minus_combinations_twist} in appendix \ref{sec:calculational_breaking_E6_to_SO10}) correspond to the irreducible representations $\boldsymbol{16} + \boldsymbol{\overline{16}}$ of $\text{SO}(10)$. Again, the details are given in appendix \ref{sec:calculational_breaking_E6_to_SO10}.

\subsection{Reducing the Rank of the Gauge Group}
\label{sec:discuss_reducing_rank}

We are now ready to consider possible Wilson lines that lead to rank reduction.
The matrix representation of the gauge twist in the the standard basis of $\mathbb{R}^8$ is
\begin{equation}
s = r_1\,r_6 = \text{diag}(1,1,1,\sigma,1,-\sigma), \quad \text{ with } \quad \sigma = 
\left(\begin{array}{cc} 0 & 1 \\ 
1 & 0 \\ 
\end{array}\right).
\label{eq:definition_of_twist}
\end{equation}
Diagonalising the matrix representation we find that there are 2 directions which are completely rotated, namely 
\begin{equation}
\lambda \, (0,\, 0,\, 0,\, 1,\, -1,\, 0,\, 0,\, 0), \quad \lambda' \, (0,\, 0,\, 0,\, 0,\, 0,\, 0,\, 1,\, 1), \quad \lambda, \lambda' \in \mathbb{R},
\label{eq:3_continuous_wilsonlines}
\end{equation}
which can be switched on as continuous Wilson lines. As we can read off immediately, these Wilson lines are
in the direction of the broken Cartan generators $H_4-H_5$ and $H_7+H_8$ of the original 
$\text{E}_6$ algebra. Switching on these Wilson lines leads to a symmetry breaking pattern as
summarised in table 2. The reduction of the rank of the gauge group is clearly demonstrated. 
We are now ready to use this result in a framework of a more realistic model.

\begin{table}[h!]
\renewcommand{\arraystretch}{1.5}
\normalsize
\begin{center}
\begin{tabular}{|l|l|l|}
\hline
First Wilson Line & Second Wilson Line & Unbroken Gauge Group\\
\hline
\hline
$\lambda \langle H_4-H_5 \rangle$	&		&	 $\text{SU}(5)\times\text{U}(1)$\\
\hline
$\lambda \langle H_7+H_8\rangle$	&	 	&	 $\text{SU}(5)\times\text{U}(1)$\\
\hline
$\lambda \langle H_4-H_5\rangle + \lambda \langle H_7+H_8 \rangle$	&	 	&	 $\text{SO}(7)\times\text{U}(1)$\\
\hline
$\lambda \langle H_4-H_5\rangle$		& $\lambda' \langle H_7+H_8 \rangle$	& $\text{SU}(4)\times\text{U}(1)$\\
\hline
$\lambda \langle H_4-H_5\rangle$		& $\lambda' \langle H_4-H_5\rangle + \lambda' \langle H_7+H_8 \rangle$	& $\text{SU}(4)\times\text{U}(1)$\\
\hline
$\lambda \langle H_7+H_8\rangle$		& $\lambda' \langle H_4-H_5\rangle + \lambda' \langle H_7+H_8 \rangle$	& $\text{SU}(4)\times\text{U}(1)$\\
\hline
\end{tabular}
\end{center}
\caption{All symmetry breakings of the type $\text{E}_6 \overset{s}{\rightarrow} \text{SO}(10)\times \text{U}(1) \overset{W}{\rightarrow} \mathfrak{g}$ with continuous Wilson lines $W$.}
\label{tab:all_possible_continuous_wilsonline_for_z2_twist}
\end{table}

\section{A $\boldsymbol{\mathbb{Z}_2}$ Orbifold in 6 Dimensions}

We shall now try to see how the technology developed so far can be implemented in the
framework of (semi) realistic model building. We envisage a situation where we start with
gauge symmetry E$_6$ in the bulk, broken by continuous and discrete Wilson lines to
the standard model gauge group. 
We shall consider the symmetry breakings discussed above in the context 
of a 6-dimensional orbifold model. Because the embedding of the point group 
in the gauge degrees of freedom $P \hookrightarrow G$ is a homomorphism, the 
order of the space-time twist is required to be a multiple the order of the root lattice automorphism, 
which happens to be 2 in our example. In the 
following, we shall therefore consider the simplest case, the $\mathbb{Z}_2$ orbifold.

\begin{figure}[h!]
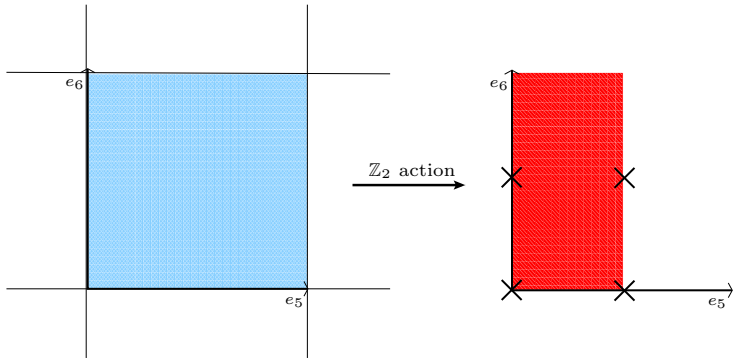

\centering
\begin{center}\input z2_orbifold.pstex_t \end{center}
\caption{The geometry of the $\mathbb{Z}_2$ orbifold.}
\label{fig:z2orbifold}
\end{figure}

The action of the translation group $L$, and of the point group $P$ are illustrated in fig.~\ref{fig:z2orbifold}. The first picture shows how the torus is defined by identifying points which differ by lattice shifts: $x \sim x + n e_5 + m e_6$, $n,m \in \mathbb{Z}$. In the second picture, the action of the point group $P$ identifies points on the torus, $x \sim -x$, and we obtain the {\it fundamental domain} of the orbifold. The 4 special points on the torus which are mapped onto themselves by the action of $P$ are called {\it fixed points}.

With the twist in the space-time we associate $s = r_1 r_6$ as the 
twist in the gauge degrees of freedom, 
as discussed in the last section. This would lead us to the gauge group
$\text{SO}(10)\times \text{U}(1)$.  
The lattice vectors $e_5$ and $e_6$ defining the torus can be 
embedded nontrivially as Wilson lines and break the gauge symmetry further. Our previous analysis
makes it clear that the breakdown to the standard model can not be achieved by just one
continuous Wilson line. Thus we need a discrete Wilson line as well. We choose
\begin{equation}
W_5 = (1/2,\, 1/2,\, -1,\, 0,\, 0,\, 0,\, -1/2,\, 1/2), \quad W_6 =
\lambda \, (0,\, 0,\, 0,\, 1,\, -1,\, 0,\, 0,\, 0), 
\label{eq:choice_of_wilsonlines}
\end{equation}
where $W_5$ represents the discrete Wilson line, while
$W_6$ corresponds to a  
continuous Wilson line in the $H_4-H_5$ direction as discussed
previously. The choice of $W_5$ is motivated by the desire to
obtain a realistic gauge group. A more detailed discussion will be given
in section \ref{sec:string}.
In the following we shall now
exhibit the ``gauge group geography'' \cite{Forste:2004ie}
in the 2-dimensional orbifold. The result is displayed in fig.2.
In the bulk we have the gauge group $\text{E}_6$. At the fixed point $(0,0)$ the
gauge symmetry is only affected by the twist and not by the Wilson lines, 
thus $\text{SO}(10)\times \text{U}(1)$.

\subsection{The symmetry at the nontrivial fixed points}
\label{sec:discrete_wilsonline}

Before discussing our concrete E$_6$ model it turns out to be useful to have
a look at the projection conditions in a general situation. A non
trivial fixed point is associated to a space group element
$\left(\theta , l\right)$ which means that a  180$^\circ$ rotation
($\theta$) is followed by a shift with the lattice vector $l$. The
non trivial fixed points on our torus correspond to space group
elements with $l = e_5, e_6,$ or  
$e_5 + e_6$ (see fig.\ \ref{fig:z2orbifold}). We want to discuss the
embedding of the space group element separately for Cartan directions
of the bulk group, step operators belonging to invariant root vectors
and step operators belonging to non invariant roots.

The transformation rule for Cartan generators is not sensitive to
Wilson lines. The projections are the same at all the fixed points. 

If the embedding of the root lattice automorphism into the algebra is
not degenerate, step operators belonging to invariant roots are
invariant under the rotation \cite{Schellekens:1987ij}. Those
operators are sensitive only to the Wilson line. They transform as
\begin{equation}\label{eq:invwilsonface}
E_{\alpha} \to e^{2\pi i \alpha\cdot W_l}E_{\alpha} ,
\end{equation}
where $W_l$ is the Wilson line on the non contractible cycle spanned
by the lattice vector $l$. If that happens to be a continuous Wilson
line, even the remaining phase is trivial because the
scalar product of an invariant root and a continuous Wilson line
vanishes. The orbifold acts on the root lattice as a rotation and
hence
$$\alpha \cdot W_l = s \alpha \cdot s W_l = -\alpha \cdot W_l , $$
where in the last step we have used that $\alpha$ is invariant and a
continuous Wilson line points into an odd Cartan direction.
The situation is different for discrete Wilson lines. The scalar
product of a discrete Wilson line with an invariant root can be half
integer and the corresponding step operator is projected out.
 
In addition we need the transformation properties of step operators
belonging to non invariant roots:
\begin{equation}\label{eq:wilsonface}
E_\alpha \to e^{2 \pi i s\alpha \cdot
  W_l}\tilde{s}E_\alpha\tilde{s}^{-1} 
\end{equation}
(recall that we first rotate and then shift). Since for non invariant
roots the orbifold image is a different step operator, the sum of
algebra element and orbifold image never vanishes for those roots.
The number of invariant sums is not altered by $W_l$. What is changed
is the way these invariant sums are embedded into the bulk gauge
algebra. In particular with a continuous Wilson line one can
{\it continuously rotate} the embedding.

These observations give an appealing geometric picture for the
symmetry breakdown by a continuous Wilson line. If the degeneracy of
two fixed points is lifted by a continuous Wilson line our above
arguments imply that the unbroken gauge groups are still the same
at these fixed points. The lifted degeneracy shows up in a misaligned
embedding of these unbroken gauge groups into the bulk group. This
results in a smaller overlap of the gauge groups at the two fixed
points and hence yields a reduced gauge symmetry in four
dimensions. The overlap does not contain invariant combinations
coming from step operators with non invariant roots and a non trivial
phase under eq.~(\ref{eq:wilsonface}).

Our geometric understanding of the symmetry breakdown with a
continuous Wilson line yields also a consistent picture in the limit
of vanishing Wilson line. Geometrically, this is the limit where the
alignment in the embedding of the gauge groups at two different fixed
points is restored. 

\begin{figure}[h!]
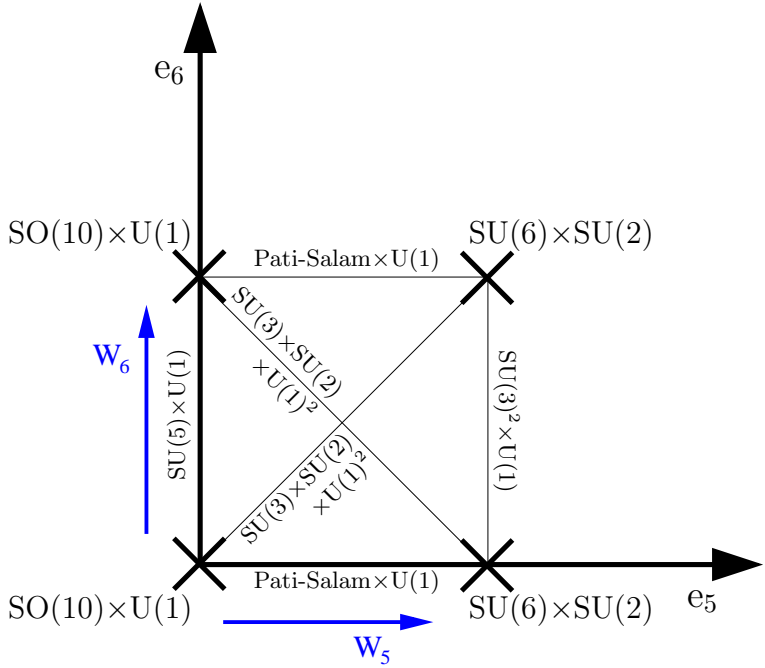

\begin{center}\input gaugegroup_geography.pstex_t \end{center}
\caption{The unbroken gauge groups at the 4 fixed points. The gauge group in the bulk is $\text{E}_6$.
A detailed explanation is given in the text.}
\label{fig:gaugegroupgeography}
\end{figure}

Let us now give the detailed picture for our particular E$_6$ model.
The gauge symmetry at the fixed point $(1/2,0)$  is sensitive to
the Wilson line 
\begin{equation}
W_5 = (1/2,\, 1/2,\, -1,\, 0,\, 0,\, 0,\, -1/2,\, 1/2),
\end{equation}
which is invariant under $s$, and hence discrete. According to eq.\
(\ref{eq:invwilsonface})  eight step operators to the $s$-invariant
spinorial roots $\beta_{57\mbox{--}60}$, $\beta_{66\mbox{--}69}$ (see
table \ref{tab:roots_E6} in the appendix) are projected out. The
bulk symmetry is broken to SU(6)$\times$SU(2) at the fixed point
$(1/2,0)$: 
\begin{equation}
\text{E}_6\rightarrow \text{SU}(6)\times\text{SU}(2), \qquad
\boldsymbol{78} \rightarrow (\boldsymbol{35}, \boldsymbol{1}) +
(\boldsymbol{1}, \boldsymbol{3}) + 
(\boldsymbol{20}, \boldsymbol{2}) .
\end{equation}
The Wilson line along $e_6$
\begin{equation}
W_6 = \lambda \, (0,\, 0,\, 0,\, 1,\, -1,\, 0,\, 0,\, 0),\quad
\lambda \in \mathbb{R}, 
\end{equation}
is completely rotated, and hence continuous. According to our general
discussion in the beginning of the section, the unbroken gauge
group for fixed points differing only in that direction is the same.
The situation is summarised in figure \ref{fig:gaugegroupgeography}
where so far we have discussed the gauge symmetry breaking at the
fixed points. 

In order to visualise the fact that the groups unbroken
at the fixed points are embedded differently into the bulk 
gauge group we have displayed
 the overlapping gauge group at lines connecting different
fixed points. Technically, these groups are obtained by imposing two
projection conditions, one for each fixed point involved. For example
at the line connecting the origin with the fixed point 
$(0,1/2)$ only those operators which are invariant
under the action of $\tilde{s}$, and $W_6$, will survive: 
\begin{equation}
E_\alpha \enspace \overset{\tilde{s}}{\rightarrow} \enspace E'_{\beta} \equiv \tilde{s}
E_\alpha \tilde{s}^{-1}, \qquad  E'_{\beta} \overset{W_6}{\rightarrow} \enspace
\exp\left(2\pi i \beta\cdot W_6 \right) E'_\beta . 
\end{equation}
The continuous Wilson line $W_6$ has the effect of projecting out all 
operators whose commutator with $H_4 - H_5$ is non-zero. The 25 surviving 
operators correspond to the unbroken gauge group
$\text{SU}(5)\times\text{U}(1)$, 
as explained in the last section.
The groups written at the other lines in fig.\
\ref{fig:gaugegroupgeography} are computed in an analogous way.

\subsection{The Spectrum in Four Dimensions}
\label{sec:first_rotate_then_wilsonline}

The unbroken gauge group in four dimensions consists of those operators which are invariant under all the symmetry transformations:
\begin{equation}
E_\alpha \enspace \overset{\tilde{s}}{\rightarrow} \enspace E'_\beta
\equiv \tilde{s} E_\alpha \tilde{s}^{-1}, \quad E'_\beta
\overset{W_5}{\rightarrow} \enspace \exp\left(2\pi i \beta\cdot W_5
\right) \, E'_\beta, \quad E''_\gamma \overset{W_6}{\rightarrow}
\enspace \exp\left(2\pi i \gamma\cdot W_6 \right) \, E''_\gamma .
\end{equation}
Alternatively, we can say that the symmetry in four dimensions is the intersection of the gauge groups at the fixed points.

In fact we would like to discuss the symmetry breaking pattern in two steps. The Wilson line $W_5$ is
discrete and its value will therefore be of the order of the string scale. $W_6$ is continuous and
will be assumed to have a smaller vacuum expectation value that breaks the remaining gauge group via
a Higgs mechanism. In the first step we then have a resulting gauge symmetry from the overlap of the
gauge symmetries at the fixed points $(0,0)$ and $(1/2,0)$. Thus the first step amounts to

\begin{equation}
 \text{SO}(10)\times\text{U(1)} \overset{W_5}{\rightarrow} 
\text{SU}(4)\times\text{SU}(2)\times\text{SU}(2)\times\text{U}(1).
\end{equation}
At the intermediate scale we thus obtain the Pati-Salam gauge group (with an additional $\text{U}(1)$ factor).
The details of the calculation can be found in appendix \ref{sec:calculational_4d_spectrum}.

In the next step we then have the breakdown of $\text{SU}(4)\times\text{SU}(2)\times\text{SU}(2)$
due to the action of the continuous Wilson line $W_6$:
\begin{equation}
\text{SU}(4)\times\text{SU}(2)\times\text{SU}(2) \overset{W_6}{\rightarrow} 
\text{SU}(3)\times\text{SU}(2)\times\text{U}(1).
\end{equation}
The continuous Wilson line thus provides a smooth breakdown of the Pati-Salam gauge group
to the standard model. From the low energy effective field theory point of view this 
corresponds to a Higgs mechanism.

The full chain of symmetry breakdown is 
\begin{equation}
\text{E}_6 \overset{\tilde{s}}{\rightarrow} \text{SO}(10)\times\text{U(1)} \overset{W_5}{\rightarrow} \text{SU}(4)\times\text{SU}(2)\times\text{SU}(2)\times\text{U}(1) \overset{W_6}{\rightarrow} \text{SU}(3)\times\text{SU}(2)\times\text{U}(1)^2.
\end{equation}
The details of the calculation are given in appendix \ref{sec:calculational_4d_spectrum}.

\section{Possible relations to string theory constructions}\label{sec:string}

So far, we have concentrated on a field theoretic orbifold model in $d=6$. Of course, this is just
a first step to understand the mechanism of rank reduction. The final aim would be to implement
the scheme in the framework of a consistent string orbifold construction. This would assure
the quantum consistency of the theory and it would also give us a hint about the incorporation
and location of matter fields. The example discussed in the last section should be used as
a tool to implement the mechanism in $d=10$ string orbifolds. To find such applications, let us 
look at some string constructions of models with Pati-Salam gauge
symmetry, as e.g.\ given 
in ref.~\cite{Kobayashi:2004ya}. The model A1 of this paper seems to be particularly suited
to our discussion. It is the result of a ${\mathbb Z}_6$ orbifold of
the $\text{E}_8\times \text{E}_8$ heterotic 
string with 3 families of quarks and leptons and a Pati-Salam (PS) gauge group. We would
like to see whether our mechanism can be applied to such a model by 
providing a smooth breakdown of the PS gauge group. 

Let us start our discussion using the notation of ref.~\cite{Kobayashi:2004ya}
where the authors describe their model in a certain
approximation as a 
 5d orbifold GUT with an $\text{E}_6$ gauge symmetry and point group $\mathbb{Z}_2$. In the bulk, there are the gauge fields in the adjoint representation $\boldsymbol{78}$, and $4 \times (\boldsymbol{27} + \boldsymbol{\overline{27}})$ hypermultiplets.

\begin{figure}[h!]
\centering
\epsfig{figure=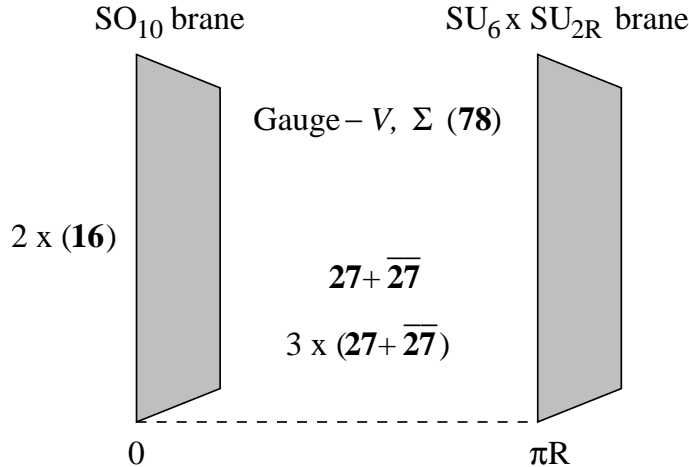, angle=0, scale=0.7}
\caption{Setup of the 5d orbifold GUT with an $\text{E}_6$ gauge symmetry of ref.~\cite{Kobayashi:2004ya}.}
\label{fig:setup_in_5d}
\end{figure}

The 2 orbifold parities break the bulk gauge group $\text{E}_6$ to $\text{SO}(10)$ at the $y=0$ brane, and to $\text{SU}(6)\times\text{SU}(2)$ at the $y=\pi R$ brane, where $y$ denotes the coordinate of the extra dimension. The setup of the model is summarised in fig.~\ref{fig:setup_in_5d}.
The gauge group in 4 dimensions is realised as the intersection of the symmetries at the 2 branes, and yields $\text{SU}(4)\times\text{SU}(2)\times\text{SU}(2)$. This Pati-Salam symmetry 
should be spontaneously broken to the Standard Model gauge group via a Higgs mechanism.

As the model of \cite{Kobayashi:2004ya} has a string theory origin in $d=10$
it naturally allows its interpretation as 
a six dimensional orbifold GUT model. The $d=6$
model is  obtained from ten dimensional heterotic string theory by
compactifying 
on $T^4/{\mathbb Z}_3$ with a discrete (third order) Wilson line. The
orbifold group in the string model is ${\mathbb Z}_6 = {\mathbb Z}_3
\times {\mathbb Z}_2$. The ${\mathbb Z}_2$ factor acts on the
remaining $T^2$ exactly in the same way as in our orbifold GUT. 
For details see 
\cite{Kobayashi:2004ya}. 
The string model yields also E$_6$ gauge symmetry in the bulk of
$T^2/{\mathbb Z}_2$ but there are additional U(1) and hidden sector
gauge group factors. Moreover, there is bulk matter coming from
twisted and untwisted sectors in the $T^4/{\mathbb Z}_3$ compactified
heterotic model. Finally, the ${\mathbb Z}_2$ twisted sector gives
information on the localisation of matter at the $T^2/{\mathbb Z}_2$
fixed points. 

Our discrete Wilson line $W_5$ is equivalent to the second
order Wilson line of ref. \cite{Kobayashi:2004ya} 
 which lifts the degeneracy of the left and right plane in figure 
\ref{fig:setup_in_5d}. 
As long as we do not switch on a continuous Wilson line, the gauge
group geography in our model and the one derived from the string model
are identical when we focus on E$_6$. 

The continuous Wilson line breaks the Pati-Salam group to the
standard model gauge group. In the low-energy effective field theory
this corresponds to a nontrivial vacuum expectation value of a
bulk field. Our analysis demonstrates that such a smooth breakdown can
be achieved within the string model considered here. This actually proves 
that there exists a bulk field in the theory that has all the 
properties of a modulus, i.e.\ a flat direction in the full scalar
potential. Without the argument given above one would have needed
to compute the full low-energy effective potential and
prove that it has a flat direction.

This is in fact a nontrivial statement. To see this let us consider
the possibility to realise the breakdown
of the the standard model gauge group SU(3)$\times$SU(2)$\times$U(1)
to SU(3)$\times$U(1): the standard Higgs mechanism of electroweak
symmetry breakdown. We still have the option of switching on the
second continuous Wilson line of section 3 along the $H_7+H_8$
direction. Unfortunately, it leads to a breakdown of
SU(3)$\times$SU(2)$\times$U(1) to SU(2)$\times$SU(2)$\times$U(1).
This shows that in the model under
consideration, the Higgs field of the standard model  cannot
correspond to a bulk field. If the standard model Higgs mechanism could
be achieved within the present framework, the Higgs boson would have to be
localised on one of the branes. 
Our analysis thus clarifies properties
of the model which otherwise would have been difficult to explain.

\section{Conclusions and outlook}

In the present paper we have explored the possibility of reducing the
rank within orbifold GUT models by means of a continuous Wilson line. An
elegant description can be given when the orbifold is embedded as a
rotation in the root lattice of the GUT group. We have seen that
finding a consistently induced algebra automorphism requires some
effort. This effort pays off when the simplicity with which a
continuous Wilson line can be introduced is encountered.
This enabled us to obtain a simple geometric picture for the symmetry
breaking. Like a discrete Wilson line also a continuous one lifts the
degeneracy of fixed points. For a continuous Wilson line, however, the
bulk group is broken at the fixed points always to the same subgroup,
merely the degeneracy in the alignment of the embedding into the bulk
group is lifted. This fits nicely with the fact that such a Wilson
line can be continuously turned off. The embedding of an unbroken
gauge group into the bulk group can be continuously rotated, a smooth
transition from a misaligned to an aligned situation exists.
Since the four dimensional gauge group is given by the intersection of
the groups at all fixed points this corresponds to a smooth symmetry
breakdown that corresponds effectively to a Higgs mechanism.

The scale of the symmetry breaking due to a continuous Wilson line can
be varied smoothly. This allows for a hierarchy in the breaking
pattern. The scale of symmetry breakings due to orbifold projections
and discrete Wilson lines is dictated by the size of the extra
dimensions and typically high. The scale of the breaking due to the
continuous Wilson line can vary smoothly and, hence, be low. In the
discussed E$_6$ example this gave the pattern that E$_6$ is broken to
the Pati-Salam group at the compactification scale, and the final
breaking to the Standard Model gauge group could occur at lower
energies. Ultimately we would try to realise the standard
Higgs mechanism as a continuous Wilson line. For this
one could also think of models where a GUT group is broken
to the Standard Model by the orbifold and discrete Wilson lines while
the electroweak symmetry breaking is realised through a
continuous Wilson line. Within GUT orbifolds one should explore
these (probably numerous) possibilities. 

The mechanism discussed
in this paper would be a useful tool to to analyse such possibilities.
It could serve as an intermediate step towards a realisation in
full $d=10$ string theory (where at the moment we are not able
to classify all the possibilities). This would allow us to
identify potentially interesting models by means of the 
simpler mechanism before one tackles the questions of the
general embedding in string theory. For the interesting cases,
our method should then be implemented in the full $d=10$ 
theory as rotations in the E$_8\times$E$_8$ or SO(32) lattices.

\bigskip

\noindent {\bf Acknowledgements}

\noindent
We would like to thank Athanasios Dedes, J\"org Meyer, and Patrick Vaudrevange for useful discussions. A.W.\
acknowledges the kind hospitality extended towards him during a visit
of Durham university. This work was
partially supported by the European community's 6th framework
programs MRTN-CT-2004-503369 ``Quest for Unification'' and
MRTN-CT-2004-005104 ``Forces Universe''.

\clearpage
\newpage

\appendix 

\section{Determining the Phase in the Lift of Weyl Reflections}
\label{sec:calculation_of_phase}

To evaluate the expression
\begin{equation}
\tilde{r}_\beta E_\alpha \tilde{r}_\beta^{-1} = \exp\left(i\frac{\pi}{2}\left(E_{\beta} + E_{-\beta} \right) \right) E_\alpha \exp\left(-i\frac{\pi}{2}\left(E_{\beta} + E_{-\beta} \right) \right),
\label{eq:evaluate_lift_for_single_weyl_reflection}
\end{equation}
we make use of the Baker-Campbell-Hausdorff formula
\begin{equation}
e^{A} B \,e^{-A} = \sum_{m=0}^\infty \frac{1}{m!} \left[A, B \right]_m, 
\quad \left[A, B \right]_m \equiv \left[A, \left[A, \ldots, \left[A, B\right]\,\right]\,\right],
\label{eq:baker_campbell_hausdorff}
\end{equation}
and set 
\begin{equation}
A = \frac{i\pi}{2}\left(E_\beta + E_{-\beta} \right), \quad B = E_\alpha,
\end{equation} 
We extend the definition of the structure constants in the sense that $N_{\alpha, \beta} = 0$, if $\alpha+\beta \notin \Delta$, where $\Delta$ denotes the set of roots. There are two cases to be distinguished. If $\alpha \neq \pm \beta$, then
\begin{equation}
\left[E_\beta+E_{-\beta}, E_\alpha \right]_{2n} = \left(N_{\beta, \alpha}\right)^{2n} E_{\alpha}, \qquad \left[E_\beta+E_{-\beta}, E_\alpha \right]_{2n+1} = \left(N_{\beta, \alpha}\right)^{2n+1} E_{\alpha+\beta},
\end{equation}
and substituting these results in eq.~(\ref{eq:evaluate_lift_for_single_weyl_reflection}), we arrive at
\begin{equation}
\tilde{r}_\beta E_\alpha \tilde{r}_\beta^{-1} = \cos\left( \frac{\pi}{2} N_{\beta, \alpha}\right) E_\alpha + i\sin\left(\frac{\pi}{2} N_{\beta, \alpha} \right) E_{\alpha+\beta}.
\end{equation}
For algebras of type $ADE$, the non-vanishing structure constants are $\pm1$. In this case, the last equation simplifies to
\begin{equation}
\tilde{r}_\beta E_\alpha \tilde{r}_\beta^{-1} = i N_{\beta, \alpha} E_{\alpha+\beta}, \quad \alpha \neq \pm \beta, \quad  N_{\beta, \alpha}\neq 0.
\label{eq:transformation_of_E_alpha_for_Nab_not_zero_simplified}
\end{equation}

In the case $\alpha = \beta$, we have
\begin{equation} 
\left[ E_\alpha + E_{-\alpha}, E_\alpha\right]_{2n} = n |\alpha|^{2n} \left( E_\alpha - E_{-\alpha} \right), \qquad \left[ E_\alpha + E_{-\alpha}, E_\alpha\right]_{2n+1} = -2n|\alpha|^{2n} \, \alpha \cdot H,
\end{equation}
and eq.~(\ref{eq:evaluate_lift_for_single_weyl_reflection}) yields
\begin{equation} 
E_\alpha - \frac{1}{2}\left( E_\alpha - E_{-\alpha}\right) + \frac{1}{2}\left(E_\alpha - E_{-\alpha} \right) \cos \pi + \frac{i}{2} \, \alpha \cdot H \left( \sin \pi \right) = E_{-\alpha}.
\end{equation} 
The result for the case $\alpha = -\beta$ is easily deduced from the previous one. In summary, we have:
\begin{equation} 
\renewcommand{\arraystretch}{1.5}
\begin{array}{rll}
\beta+\alpha \in \Delta, \quad \beta \neq \pm \alpha \quad &\rightarrow& \quad \tilde{r}_\alpha E_\beta \tilde{r}_\alpha^{-1} = i N_{\alpha, \beta} E_{\beta+\alpha}\\
\beta-\alpha \in \Delta, \quad \beta \neq \pm \alpha \quad &\rightarrow& \quad \tilde{r}_\alpha E_\beta \tilde{r}_\alpha^{-1} = i N_{-\alpha, \beta} E_{\beta-\alpha}\\
\beta\pm\alpha \notin \Delta, \quad \beta \neq \pm \alpha \quad &\rightarrow& \quad \tilde{r}_\alpha E_\beta \tilde{r}_\alpha^{-1} = E_\beta\\
\beta = \pm\alpha \quad &\rightarrow& \quad \tilde{r}_\alpha E_\beta \tilde{r}_\alpha^{-1} = E_{-\beta}
\end{array}
\label{eq:transformation_properties_of_all_step_operators}
\end{equation}

The determination of the structure constants is described in
ref.~\cite{Carter:1972wk}. 


\section{Choice of roots and simple roots for $\boldsymbol{\text{E}_6}$}
\label{sec:choice_roots}

We describe $\text{E}_6$ in terms of its embedding in $\text{E}_8$. The roots of $\text{E}_6$ are those roots of $\text{E}_8$, whose first 3 components are equal. For convenience and reference, we list the roots in tab.~\ref{tab:roots_E6}. Using the standard metric on root space, we find the simple roots as listed in tab.~\ref{tab:simpleroots_E6}.

\begin{table}[h!]
\renewcommand{\arraystretch}{1}
\small
\begin{center}
\begin{tabular}{|l|rrrrrrrr|}
\hline
$\alpha_{1}$ & 0 &  0 &  0 &  1 & -1 &  0 &  0 &  0 \\
$\alpha_{2}$ & 0 &  0 &  0 &  0 &  1 & -1 &  0 &  0 \\
$\alpha_{3}$ & 0 &  0 &  0 &  0 &  0 &  1 & -1 &  0 \\
$\alpha_{4}$ & 0 &  0 &  0 &  0 &  0 &  0 &  1 & -1 \\
$\alpha_{5}$ & 1/2 &  1/2 &  1/2 & -1/2 & -1/2 & -1/2 & -1/2 &  1/2 \\
$\alpha_{6}$ & 0 &  0 &  0 &  0 &  0 &  0 &  1 &  1 \\
\hline
\end{tabular}
\end{center}
\caption{Simple roots of $\text{E}_6$.}
\label{tab:simpleroots_E6}
\end{table}

\begin{table}[h!]
\renewcommand{\arraystretch}{1}
\scriptsize
\begin{center}
\begin{tabular}{|l|rrrrrrrr|l|l|rrrrrrrr|}
\cline{1-9} \cline{11-19}
$\beta_{1}$ & 0 &  0 &  0 &  1 &  1 &  0 &  0 &  0 &    & $\beta_{37}$ & 0 &  0 &  0 &  0 &  0 &  0 &  1 &  1 \\		       
$\beta_{2}$ & 0 &  0 &  0 & -1 &  1 &  0 &  0 &  0 &	& $\beta_{38}$ & 0 &  0 &  0 &  0 &  0 &  0 & -1 &  1 \\		       
$\beta_{3}$ & 0 &  0 &  0 &  1 & -1 &  0 &  0 &  0 &	& $\beta_{39}$ & 0 &  0 &  0 &  0 &  0 &  0 &  1 & -1 \\		       
$\beta_{4}$ & 0 &  0 &  0 & -1 & -1 &  0 &  0 &  0 &	& $\beta_{40}$ & 0 &  0 &  0 &  0 &  0 &  0 & -1 & -1 \\		       
$\beta_{5}$ & 0 &  0 &  0 &  1 &  0 &  1 &  0 &  0 &	& $\beta_{41}$ & 1/2 &  1/2 &  1/2 &  1/2 &  1/2 &  1/2 &  1/2 &  1/2 \\ 
$\beta_{6}$ & 0 &  0 &  0 & -1 &  0 &  1 &  0 &  0 &	& $\beta_{42}$ & -1/2 & -1/2 & -1/2 & -1/2 & -1/2 & -1/2 & -1/2 & -1/2 \\
$\beta_{7}$ & 0 &  0 &  0 &  1 &  0 & -1 &  0 &  0 &	& $\beta_{43}$ & 1/2 &  1/2 &  1/2 & -1/2 & -1/2 &  1/2 &  1/2 &  1/2 \\ 
$\beta_{8}$ & 0 &  0 &  0 & -1 &  0 & -1 &  0 &  0 &	& $\beta_{44}$ & -1/2 & -1/2 & -1/2 &  1/2 &  1/2 & -1/2 & -1/2 & -1/2 \\
$\beta_{9}$ & 0 &  0 &  0 &  1 &  0 &  0 &  1 &  0 &	& $\beta_{45}$ & 1/2 &  1/2 &  1/2 & -1/2 &  1/2 & -1/2 &  1/2 &  1/2 \\ 
$\beta_{10}$ & 0 &  0 &  0 & -1 &  0 &  0 &  1 &  0 &	& $\beta_{46}$ & -1/2 & -1/2 & -1/2 &  1/2 & -1/2 &  1/2 & -1/2 & -1/2 \\
$\beta_{11}$ & 0 &  0 &  0 &  1 &  0 &  0 & -1 &  0 &	& $\beta_{47}$ & 1/2 &  1/2 &  1/2 & -1/2 &  1/2 &  1/2 & -1/2 &  1/2 \\ 
$\beta_{12}$ & 0 &  0 &  0 & -1 &  0 &  0 & -1 &  0 &	& $\beta_{48}$ & -1/2 & -1/2 & -1/2 &  1/2 & -1/2 & -1/2 &  1/2 & -1/2 \\
$\beta_{13}$ & 0 &  0 &  0 &  1 &  0 &  0 &  0 &  1 &	& $\beta_{49}$ & 1/2 &  1/2 &  1/2 & -1/2 &  1/2 &  1/2 &  1/2 & -1/2 \\ 
$\beta_{14}$ & 0 &  0 &  0 & -1 &  0 &  0 &  0 &  1 &	& $\beta_{50}$ & -1/2 & -1/2 & -1/2 &  1/2 & -1/2 & -1/2 & -1/2 &  1/2 \\
$\beta_{15}$ & 0 &  0 &  0 &  1 &  0 &  0 &  0 & -1 &	& $\beta_{51}$ & 1/2 &  1/2 &  1/2 &  1/2 & -1/2 & -1/2 &  1/2 &  1/2 \\ 
$\beta_{16}$ & 0 &  0 &  0 & -1 &  0 &  0 &  0 & -1 &	& $\beta_{52}$ & -1/2 & -1/2 & -1/2 & -1/2 &  1/2 &  1/2 & -1/2 & -1/2 \\
$\beta_{17}$ & 0 &  0 &  0 &  0 &  1 &  1 &  0 &  0 &	& $\beta_{53}$ & 1/2 &  1/2 &  1/2 &  1/2 & -1/2 &  1/2 & -1/2 &  1/2 \\ 
$\beta_{18}$ & 0 &  0 &  0 &  0 & -1 &  1 &  0 &  0 &	& $\beta_{54}$ & -1/2 & -1/2 & -1/2 & -1/2 &  1/2 & -1/2 &  1/2 & -1/2 \\
$\beta_{19}$ & 0 &  0 &  0 &  0 &  1 & -1 &  0 &  0 &	& $\beta_{55}$ & 1/2 &  1/2 &  1/2 &  1/2 & -1/2 &  1/2 &  1/2 & -1/2 \\ 
$\beta_{20}$ & 0 &  0 &  0 &  0 & -1 & -1 &  0 &  0 &	& $\beta_{56}$ & -1/2 & -1/2 & -1/2 & -1/2 &  1/2 & -1/2 & -1/2 &  1/2 \\
$\beta_{21}$ & 0 &  0 &  0 &  0 &  1 &  0 &  1 &  0 &	& $\beta_{57}$ & 1/2 &  1/2 &  1/2 &  1/2 &  1/2 & -1/2 & -1/2 &  1/2 \\ 
$\beta_{22}$ & 0 &  0 &  0 &  0 & -1 &  0 &  1 &  0 &	& $\beta_{58}$ & -1/2 & -1/2 & -1/2 & -1/2 & -1/2 &  1/2 &  1/2 & -1/2 \\
$\beta_{23}$ & 0 &  0 &  0 &  0 &  1 &  0 & -1 &  0 &	& $\beta_{59}$ & 1/2 &  1/2 &  1/2 &  1/2 &  1/2 & -1/2 &  1/2 & -1/2 \\ 
$\beta_{24}$ & 0 &  0 &  0 &  0 & -1 &  0 & -1 &  0 &	& $\beta_{60}$ & -1/2 & -1/2 & -1/2 & -1/2 & -1/2 &  1/2 & -1/2 &  1/2 \\
$\beta_{25}$ & 0 &  0 &  0 &  0 &  1 &  0 &  0 &  1 &	& $\beta_{61}$ & 1/2 &  1/2 &  1/2 &  1/2 &  1/2 &  1/2 & -1/2 & -1/2 \\ 
$\beta_{26}$ & 0 &  0 &  0 &  0 & -1 &  0 &  0 &  1 &	& $\beta_{62}$ & -1/2 & -1/2 & -1/2 & -1/2 & -1/2 & -1/2 &  1/2 &  1/2 \\
$\beta_{27}$ & 0 &  0 &  0 &  0 &  1 &  0 &  0 & -1 &	& $\beta_{63}$ & -1/2 & -1/2 & -1/2 & -1/2 &  1/2 &  1/2 &  1/2 &  1/2 \\
$\beta_{28}$ & 0 &  0 &  0 &  0 & -1 &  0 &  0 & -1 &	& $\beta_{64}$ & -1/2 & -1/2 & -1/2 &  1/2 & -1/2 &  1/2 &  1/2 &  1/2 \\
$\beta_{29}$ & 0 &  0 &  0 &  0 &  0 &  1 &  1 &  0 &	& $\beta_{65}$ & -1/2 & -1/2 & -1/2 &  1/2 &  1/2 & -1/2 &  1/2 &  1/2 \\
$\beta_{30}$ & 0 &  0 &  0 &  0 &  0 & -1 &  1 &  0 &	& $\beta_{66}$ & -1/2 & -1/2 & -1/2 &  1/2 &  1/2 &  1/2 & -1/2 &  1/2 \\
$\beta_{31}$ & 0 &  0 &  0 &  0 &  0 &  1 & -1 &  0 &	& $\beta_{67}$ & -1/2 & -1/2 & -1/2 &  1/2 &  1/2 &  1/2 &  1/2 & -1/2 \\
$\beta_{32}$ & 0 &  0 &  0 &  0 &  0 & -1 & -1 &  0 &	& $\beta_{68}$ & 1/2 &  1/2 &  1/2 & -1/2 & -1/2 & -1/2 & -1/2 &  1/2 \\ 
$\beta_{33}$ & 0 &  0 &  0 &  0 &  0 &  1 &  0 &  1 &	& $\beta_{69}$ & 1/2 &  1/2 &  1/2 & -1/2 & -1/2 & -1/2 &  1/2 & -1/2 \\ 
$\beta_{34}$ & 0 &  0 &  0 &  0 &  0 & -1 &  0 &  1 &	& $\beta_{70}$ & 1/2 &  1/2 &  1/2 & -1/2 & -1/2 &  1/2 & -1/2 & -1/2 \\ 
$\beta_{35}$ & 0 &  0 &  0 &  0 &  0 &  1 &  0 & -1 &	& $\beta_{71}$ & 1/2 &  1/2 &  1/2 & -1/2 &  1/2 & -1/2 & -1/2 & -1/2 \\ 
$\beta_{36}$ & 0 &  0 &  0 &  0 &  0 & -1 &  0 & -1 &	& $\beta_{72}$ & 1/2 &  1/2 &  1/2 &  1/2 & -1/2 & -1/2 & -1/2 & -1/2 \\ 
\cline{1-9} \cline{11-19}
\end{tabular}
\end{center}
\caption{Roots of $\text{E}_6$.}
\label{tab:roots_E6}
\end{table}


\section{Calculational Details of Section \ref{sec:breaking_E6_to_SO10}}
\label{sec:calculational_breaking_E6_to_SO10}

In the following, we use standard techniques of group theory to identify the gauge symmetry and the irreducible representations \cite{Cahn:1985wk, Humphreys:1980dw}.

We start with the 46 invariant combinations of operators listed in tab.~\ref{tab:invariant_combinations_twist}. As one can easily verify, the operators
\begin{gather}
\tilde{H}_1 = H_1+H_2+H_3, \quad \tilde{H}_2 = H_4+H_5, \quad \tilde{H}_3 = H_6,\notag\\
\tilde{H}_4 = H_7-H_8, \quad \tilde{H}_5 = E_2+E_3, \quad \tilde{H}_6 = E_{37}+E_{40}, \quad 
\end{gather}
form the Cartan subalgebra of the unbroken gauge group. As a first step, we identify the $\text{U}(1)$ generator, which is given by a linear combination of the Cartans, commuting with all operators in the algebra. In practical terms, evaluating the Killing form on the Cartan generators, $\tilde{K}_{ij} = \text{Tr } \text{ad } \tilde{H}_i \,\, \text{ad } \tilde{H}_j$, and calculating the kernel of $\tilde{K}_{ij}$ gives the $\text{U}(1)$ generator
\begin{equation}
U = \frac{1}{16}\left( \tilde{H_1} + 3\,\tilde{H_3} + 3\,\tilde{H_5} + 3\,\tilde{H_6} \right).
\end{equation}

\begin{table}[h!]
\renewcommand{\arraystretch}{1}
\small
\begin{center}
\begin{tabular}{|l|l|l|l|l|l|l|l|l|l|l|l|}
\cline{1-1} \cline{3-3} \cline{5-5} \cline{7-7} \cline{9-9} \cline{11-11} 
  $E_{1}$ & &            $E_{10} + E_{28}$ & &   $E_{30} + E_{36}$ & &   $E_{43} + E_{70}$ & &   $E_{51} - E_{71}$ & &   $E_{68}$\\
  $E_{2} + E_{3}$ & &    $E_{11} + E_{25}$ & &   $E_{31} + E_{33}$ & &   $E_{44} + E_{65}$ & &   $E_{52} - E_{64}$ & &   $E_{69}$\\
  $E_{4}$ & & 	       $E_{12} + E_{26}$ & &   $E_{32} + E_{34}$ & &   $E_{45} - E_{72}$ & &   $E_{57}$ & &            $H_1 + H_2 + H_3$\\
  $E_{5} - E_{17}$ & &   $E_{13} + E_{23}$ & &   $E_{37} + E_{40}$ & &   $E_{46} - E_{63}$ & &   $E_{58}$ & &            $H_4 + H_5$\\
  $E_{6} + E_{18}$ & &   $E_{14} + E_{24}$ & &   $E_{38}$ & &            $E_{47} + E_{53}$ & &   $E_{59}$ & &            $H_6$\\      
  $E_{7} + E_{19}$ & &   $E_{15} + E_{21}$ & &   $E_{39}$ & &            $E_{48} + E_{54}$ & &   $E_{60}$ & &            $H_7 - H_8$\\     
  $E_{8} - E_{20}$ & &   $E_{16} + E_{22}$ & &   $E_{41} + E_{61}$ & &   $E_{49} + E_{55}$ & &   $E_{66}$ & &            \\
  $E_{9} + E_{27}$ & &   $E_{29} + E_{35}$ & &   $E_{42} + E_{62}$ & &   $E_{50} + E_{56}$ & &   $E_{67}$ & &            \\
\cline{1-1} \cline{3-3} \cline{5-5} \cline{7-7} \cline{9-9} \cline{11-11} 
\end{tabular}
\end{center}
\caption{The 46 invariant combinations corresponding to $\text{SO}(10)\times\text{U}(1)$.}
\label{tab:invariant_combinations_twist}
\end{table}

Next, we perform the Levi decomposition \cite{Fuchs:1997jv}, i.e.~we separate the $\text{U}(1)$ factor from the semisimple part of the algebra \cite{Rand:1986wh, GAP4}. The Cartan generators of the semisimple part of the unbroken gauge group are then given by
\begin{gather}
\tilde{H}'_1 = \tilde{H}_2, \quad \tilde{H}'_2 = \tilde{H}_3 - U, \quad \tilde{H}'_3 = \tilde{H}_4,\notag\\
\tilde{H}'_4 = \tilde{H}_5 - 2U, \quad \tilde{H}'_5 = \tilde{H}_6 - 2U,
\end{gather}
whereas the other 40 operators are unaffected. The Killing form of the semisimple part is
\begin{equation} 
\tilde{K}'_{ij} = \left(\begin{array}{ccccc} 
32    &     0     &    0    &     0   &      0\\
 0    &    13     &    0    &    -6   &     -6\\
 0    &     0     &   32    &     0   &      0\\
 0    &    -6     &    0    &    20   &    -12\\
 0    &    -6     &    0    &   -12   &     20\\
\end{array}\right).
\label{eq:killingform_so10}
\end{equation}

To find the roots, we calculate the adjoint action $\text{ad }\tilde{H}'_i$ of the Cartan generators on the algebra. These $45\times45$ matrices will in general not be diagonal, but since the Cartan generators commute, they can be simultaneously diagonalised. The $k$th eigenvalue of the $i$th matrix is then the $i$th entry of the $k$th root, $\tilde{\alpha}_k^{(i)}$.

We introduce a semi-ordering by fixing the basis
\begin{gather} 
(2,\,0,\,0,\,0,\,0), \quad (0,\,0,\,-2,\,0,\,0)\notag\\
(-1,\,0,\,1,\,1,\,-1), \quad (1,\,1,\,0,\,-1,\,0), \quad (0,\,-1/2,\,0,\,1,\,-1).
\end{gather}

Among the positive roots, we identify the simple roots as those which cannot be written as the sum of 2 positive ones:
\begin{gather} 
\alpha_1 = (0,\,-1/2,\,0,\,1,\,-1), \quad \alpha_2 = (1,\,1,\,0,\,-1,\,0),\notag\\
\alpha_3 = (0,\,0,\,-2,\,0,\,0),\quad \alpha_4 = (1,\,-1/2,\,0,\,0,\,1),\quad \alpha_5 = (-1,\,1/2,\,1,\,0,\,0)
\end{gather}

The canonical isomorphism between the Cartan subalgebra $\mathfrak{h}$ and its dual $\mathfrak{h}^*$ maps each simple root to an element of the Cartan subalgebra:
\begin{gather} 
h_{\alpha_1} = -\frac{1}{8}\tilde{H}'_2 - \frac{1}{16}\tilde{H}'_4 - \frac{1}{8}\tilde{H}'_5, \quad h_{\alpha_2} = \frac{1}{32}\tilde{H}'_1 + \frac{1}{16}\tilde{H}'_2 - \frac{1}{32}\tilde{H}'_4, \quad h_{\alpha_3} = -\frac{1}{16}\tilde{H}'_3\notag\\
h_{\alpha_4} = \frac{1}{32}\tilde{H}'_1 + \frac{1}{16}\tilde{H}'_2 + \frac{3}{32}\tilde{H}'_4 + \frac{1}{8}\tilde{H}'_5, \quad h_{\alpha_5} = -\frac{1}{32}\tilde{H}'_1 + \frac{1}{8}\tilde{H}'_2 + \frac{1}{32}\tilde{H}'_3  + \frac{3}{32}\tilde{H}'_4 + \frac{3}{32}\tilde{H}'_5
\end{gather} 

The scalar product for the root vectors $\alpha, \beta \in \mathfrak{h}^*$ is then defined by 
\begin{equation}
\langle \alpha, \beta \rangle \equiv \text{Tr } \text{ad } h_\alpha \,\, \text{ad } h_\beta.
\end{equation}
Using the Killing form given in eq.~(\ref{eq:killingform_so10}), we can evaluate the right-hand-side, and calculate the Cartan matrix
\begin{equation} 
A_{ij} \equiv 2\frac{\langle \alpha_i, \alpha_j \rangle}{\langle \alpha_j, \alpha_j \rangle} = \left(\begin{array}{ccccc} 
 2   &     -1    &     0   &     -1   &     -1\\
-1   &      2    &     0   &      0   &      0\\
 0   &      0    &     2   &      0   &     -1\\
-1   &      0    &     0   &      2   &      0\\
-1   &      0    &    -1   &      0   &      2\\
\end{array}\right).
\end{equation}
From the corresponding Dynkin diagram
\begin{center}
\begin{picture}(100,48)
\thicklines
\put(6, 14){\circle{8}} \put(36,14){\circle{8}} \put(66,14){\circle{8}} \put(96,14){\circle{8}} \put(66,44){\circle{8}}
\put(10,14){\line(1,0){22}} \put(40,14){\line(1,0){22}} \put(70,14){\line(1,0){22}} \put(66,18){\line(0,1){22}}
\put(2, 0){$\alpha_3$}
\put(32, 0){$\alpha_5$}
\put(62, 0){$\alpha_1$}
\put(92, 0){$\alpha_2$}
\put(74, 42){$\alpha_4$}
\end{picture}
\end{center}
we see that the semisimple part of the gauge group is $\text{SO}(10)$.

\begin{table}[h!]
\renewcommand{\arraystretch}{1}
\small
\begin{center}
\begin{tabular}{|l|l|l|l|l|l|l|l|}
\cline{1-1} \cline{3-3} \cline{5-5} \cline{7-7} 
$E_{2} - E_{3}$ & \phantom{XX}&     $E_{12} - E_{26}$ & \phantom{XX} &   $E_{32} - E_{34}$ & \phantom{XX} & $E_{47} - E_{53}$ \\
$E_{5} + E_{17}$ & &    $E_{13} - E_{23}$ & &   $E_{37} - E_{40}$ & &   $E_{48} - E_{54}$ \\ 
$E_{6} - E_{18}$ & &    $E_{14} - E_{24}$ & &   $E_{41} - E_{61}$ & &   $E_{49} - E_{55}$ \\ 
$E_{7} - E_{19}$ & &    $E_{15} - E_{21}$ & &   $E_{42} - E_{62}$ & &   $E_{50} - E_{56}$ \\ 
$E_{8} + E_{20}$ & &    $E_{16} - E_{22}$ & &   $E_{43} - E_{70}$ & &   $E_{51} + E_{71}$ \\ 
$E_{9} - E_{27}$ & &    $E_{29} - E_{35}$ & &   $E_{44} - E_{65}$ & &   $E_{52} + E_{64}$ \\ 
$E_{10} - E_{28}$ & &   $E_{30} - E_{36}$ & &   $E_{45} + E_{72}$ & &   $H_{4} - H_{5}$   \\ 
$E_{11} - E_{25}$ & &   $E_{31} - E_{33}$ & &   $E_{46} + E_{63}$ & &   $H_{7} + H_{8}$   \\ 
\cline{1-1} \cline{3-3} \cline{5-5} \cline{7-7}
\end{tabular}
\end{center}
\caption{The 32 combinations which transform with a minus.}
\label{tab:minus_combinations_twist}
\end{table}      

Now consider the 32 operators in tab.~\ref{tab:minus_combinations_twist} transforming with a minus sign. To calculate their weight vectors is completely analogous to the case of the adjoint representation. First, we determine the adjoint action of the 5 Cartan generators on these operators, which is given by $32\times32$ matrices. Second, these matrices are simultaneously diagonalised, and the 32 eigenvalues in the 5 diagonal matrices constitute the 32 weight vectors. Third, we take the scalar products of the weights with the simple roots to calculate the Dynkin labels. Looking for the highest weights, we find 
\begin{equation}
(0\,0 \,0\,1\,0) \text{ and } (0\,0\,0\,0\,1),
\end{equation}
corresponding to the representations $\boldsymbol{16}$ and $\boldsymbol{\overline{16}}$.


\section{Calculational Details of Section \ref{sec:first_rotate_then_wilsonline}}
\label{sec:calculational_4d_spectrum}

We first give a short description of how $\text{E}_6$ breaks to Pati-Salam $\times$ $\text{U}(1)$. Consider
\begin{equation}
E_\alpha \enspace \overset{\tilde{s}}{\rightarrow} \enspace E'_\beta \equiv \tilde{s} E_\alpha \tilde{s}^{-1}, \qquad E'_\beta \overset{W_5}{\rightarrow} \enspace \exp\left(2\pi i \beta\cdot W_5 \right) \, E'_\beta.
\end{equation}
The first transformation breaks $\text{E}_6$ to
$\text{SO}(10)\times\text{U}(1)$, see appendix
\ref{sec:calculational_breaking_E6_to_SO10}. In
tab.~\ref{tab:invariant_combinations_4d}, we list the invariant
combinations of the 
$\text{SO}(10)\times\text{U}(1)$ operators under the Wilson line
$W_5$. 

\bigskip

\begin{table}[h!]
\renewcommand{\arraystretch}{1.3}
\normalsize
\begin{center}
\begin{tabular}{|l|l|l|l|l|l|l|l|}
\cline{1-1} \cline{3-3} \cline{5-5} \cline{7-7}
$E_{1}$ & \phantom{X} & $E_{8} - E_{20}$ & \phantom{X} &   $E_{43} + E_{70}$ & \phantom{X} &  $H_{1}+H_{2}+H_{3}$\\
$E_{2} + E_{3}$ & &   $E_{37} + E_{40}$ & &  $E_{44} + E_{65}$ & &  $H_{4}+H_{5}$ \\
$E_{4}$ & &  	   $E_{38}$ & &  	       $E_{45} - E_{72}$ & &  $H_{6}$ \\  
$E_{5} - E_{17}$ & &  $E_{39}$ & &  	       $E_{46} - E_{63}$ & &  $H_{7}-H_{8}$ \\         
$E_{6} + E_{18}$ & &  $E_{41} + E_{61}$ & &  $E_{51} - E_{71}$ & & \\
$E_{7} + E_{19}$ & &  $E_{42} + E_{62}$ & &  $E_{52} - E_{64}$ & & \\
\cline{1-1} \cline{3-3} \cline{5-5} \cline{7-7}
\end{tabular}
\end{center}
\caption{The 22 invariant combinations corresponding to $\text{SU}(4)\times\text{SU}(2)\times\text{SU}(2)\times\text{U}(1)$.}
\label{tab:invariant_combinations_4d}
\end{table}

\bigskip

The 24 combinations of operators which transform with a minus sign are given in tab.~\ref{tab:matter_4d}. Calculating the Dynkin labels, we identify the highest weight $(0 1 0 | 1 | 1)$ corresponding to the representation $(\boldsymbol{6}, \boldsymbol{2}, \boldsymbol{2})$.

\bigskip

\begin{table}[h!]
\renewcommand{\arraystretch}{1.3}
\normalsize
\begin{center}
\begin{tabular}{|l|l|l|l|l|l|l|l|}
\cline{1-1} \cline{3-3} \cline{5-5} \cline{7-7}
$E_{9} + E_{27}$   & \phantom{XX} &  $E_{15} + E_{21}$  &\phantom{XX} &  $E_{47} + E_{53}$ 	 &\phantom{XX} & 	$E_{59}$\\
$E_{10} + E_{28}$  & &  $E_{16} + E_{22}$  & &  $E_{48} +E_{54}$  	 & & 	$E_{60}$\\
$E_{11} + E_{25}$  & &  $E_{29} + E_{35}$  & &  $E_{49} +E_{55}$  	 & & 	$E_{66}$\\
$E_{12} + E_{26}$  & &  $E_{30} + E_{36}$  & &  $E_{50} +E_{56}$  	 & & 	$E_{67}$\\
$E_{13} + E_{23}$  & &  $E_{31} + E_{33}$  & &  $E_{57}$	    	 & & 	$E_{68}$\\
$E_{14} + E_{24}$  & &  $E_{32} + E_{34}$  & &  $E_{58}$        	 & & 	$E_{69}$\\
\cline{1-1} \cline{3-3} \cline{5-5} \cline{7-7}
\end{tabular}
\end{center}
\caption{The 24 combinations corresponding transforming with a minus sign.}
\label{tab:matter_4d}
\end{table}

\begin{table}[h!]
\renewcommand{\arraystretch}{1.2}
\small
\begin{center}
\begin{tabular}{|l|l|l|l|l|l|l|l|l|l|}
\cline{1-1} \cline{3-3} \cline{5-5} \cline{7-7} \cline{9-9}
$E_{1}$  & \phantom{X} &            $E_{38}$  & \phantom{X}  &  	        $E_{42} + E_{62}$  & &  $H_{1}+H_{2}+H_{3}$  & \phantom{X}  &   $H_{7}-H_{8}$ \\
$E_{4}$  & &  	    $E_{39}$  & &  	    	$E_{43} + E_{70}$  & &  $H_{4}+H_{5}$  & &  	                \\
$E_{37} + E_{40}$  & &  $E_{41} + E_{61}$  & &  $E_{44} + E_{65}$  & &  $H_{6}$  & &                            \\
\cline{1-1} \cline{3-3} \cline{5-5} \cline{7-7} \cline{9-9}
\end{tabular}
\end{center}
\caption{The 13 invariant combinations corresponding to $\text{SU}(3)\times\text{SU}(2)\times\text{U}(1)\times\text{U}(1)$.}
\label{tab:invariant_combinations_4d_after_continuous_wl}
\end{table}

Next, we give a detailed description of how the Pati-Salam gauge group breaks down to the Standard Model gauge symmetry. 

The continuous Wilson line $W_6$ will project out all those step operators in tab.~\ref{tab:invariant_combinations_4d}, which have a non-vanishing scalar product with $H_4 - H_5$. Equivalently we can say that a step operator is projected out, if the corresponding root has a non-vanishing scalar product with $W_6$. The surviving operators are listed in tab.~\ref{tab:invariant_combinations_4d_after_continuous_wl}.

The Cartan subalgebra is given by
\begin{gather}
\bar{H}_1 = E_{37}+E_{40}, \quad \bar{H}_2 = H_1+H_2+H_3, \notag\\
\bar{H}_3 = H_4+H_5, \quad \bar{H}_4 = H_6, \quad \bar{H}_5 = H_7-H_8.
\end{gather}
Calculating the kernel of the Killing form 
\begin{equation}
\bar{K}_{ij} \equiv \text{Tr } \text{ad } \bar{H}_i \,\, \text{ad } \bar{H}_j = 
\left(\begin{array}{ccccc} 
  4     &   -6    &     0    &    -2    &     0\\ 
 -6     &    9    &     0    &     3    &     0\\ 
  0     &    0    &    12    &     0    &     0\\ 
 -2     &    3    &     0    &     1    &     0\\ 
  0     &    0    &     0    &     0     &    8\\ 
\end{array}\right),
\end{equation}
we find the $\text{U}(1)$ generators,
\begin{equation}
\bar{U}_1 = \bar{H}_2 - 3 \bar{H}_4, \quad \bar{U}_2 = \bar{H}_1 + 2 \bar{H}_4,
\label{eq:u1charges_SM}
\end{equation}
and after the Levi decomposition, the Cartan generators of the semisimple part of the algebra are
\begin{equation}
\bar{H}'_1 = \bar{H}_3, \quad \bar{H}'_2 = \bar{H}_4 + \frac{1}{6}\bar{U}_1 - \frac{1}{6}\bar{U}_2, \quad \bar{H}'_3 = \bar{H}_5.
\end{equation}
The Killing form of the semisimple part is then
\begin{equation}
\bar{K}'_{ij} \equiv \text{Tr } \text{ad } \bar{H}'_i \,\, \text{ad } \bar{H}'_j = 
\left(\begin{array}{ccc} 
12    &     0    &     0 \\
 0    &     1    &     0 \\
 0    &     0    &     8 \\
\end{array}\right).
\label{eq:killingform_SM}
\end{equation}

The adjoint action $\text{ad } \bar{H}'_i$ of the Cartan generators on the 11 operators of the semisimple part of the algebra,
\begin{center}
\begin{tabular}{llllll}
$E_{1}$, & $E_{38}$, & $E_{41} + E_{61}$, & $E_{43} + E_{70}$,  &     $\bar{H}'_{1}$, & $\bar{H}'_{3}$,  \\
$E_{4}$, & $E_{39}$, & $E_{42} + E_{62}$, & $E_{44} + E_{65}$,  &	  $\bar{H}'_{2}$,  &                     \\  
\end{tabular}
\end{center}
is given by $11\times11$ matrices. Since the Cartan generators mutually commute, these matrices can be simultaneously diagonalised. The $k$th eigenvalue of the $i$th matrix is then the $i$th entry of the $k$th root, $\bar{\alpha}_k^{(i)}$:
\begin{center}
\begin{tabular}{llllll}
(2,\,0 ,\,0), &  (0,\,0,\,2),    & (-1,\, 1/2,\,0), &   (0,\,0,\,-2),   &  (0,\,0,\,0), & (0,\,0,\,0),\\
(-2,\,0,\,0), &  (1,\,1/2,\,0), & (1,\,-1/2,\,0),  &   (-1,\,-1/2,\,0)	&  (0,\,0,\,0) & \\
\end{tabular}
\end{center}

We introduce a semi-ordering by fixing the basis
\begin{equation} 
(2,\,0,\,0), \quad (0,\,0,\,-2), \quad (1,\,1/2,\,0).
\end{equation}
Among the positive roots,
\begin{equation} 
(2,\,0,\,0), \quad (0,\,0,\,-2), \quad (1,\,1/2,\,0), \quad (1,\,-1/2,\,0),
\end{equation} 
we select those, which cannot be written as the sum of two positive ones ({\it simple roots}):
\begin{equation} 
\alpha_1 = (0,\,0,\,-2), \quad \alpha_2 = (1,\,1/2,\,0), \quad \alpha_3 = (1,\,-1/2,\,0).
\end{equation}
The canonical isomorphism between the Cartan subalgebra $\mathfrak{h}$, and it dual $\mathfrak{h}^*$ assigns to each simple root $\alpha_i$ an element of the Cartan subalgebra $h_{\alpha_i}$, 
\begin{equation} 
h_{\alpha_1} = -\frac{1}{4}\bar{H}'_3, \quad h_{\alpha_2} = \frac{1}{12}\bar{H}'_1 + \frac{1}{2}\bar{H}'_2, \quad h_{\alpha_3} = \frac{1}{12}\bar{H}'_1 - \frac{1}{2}\bar{H}'_2.
\end{equation} 
The scalar product in root space is then given by
\begin{equation}
\langle \alpha_i, \alpha_j \rangle \equiv \text{Tr } \text{ad } h_{\alpha_i} \,\, \text{ad } h_{\alpha_j}.
\end{equation}
Using the Killing form (cf.~eq.~(\ref{eq:killingform_SM})), the right-hand-side can easily be evaluated. From the Cartan matrix
\begin{equation} 
A_{ij} \equiv 2\frac{\langle \alpha_i, \alpha_j \rangle}{\langle \alpha_j, \alpha_j \rangle} = \left(\begin{array}{ccc} 
2    &     0    &     0 \\
0    &     2    &     -1 \\
0    &     -1    &     2 \\
\end{array}\right),
\end{equation}
and its corresponding Dynkin diagram
\begin{center}
\begin{picture}(78,18)
\thicklines
\put(6, 14){\circle{8}} \put(46,14){\circle{8}} \put(76,14){\circle{8}}
\put(50,14){\line(1,0){22}}
\put(2, 0){$\alpha_1$}
\put(42, 0){$\alpha_2$}
\put(72, 0){$\alpha_3$}
\end{picture}
\end{center}
we see that the semisimple part of the gauge group is $\text{SU}(3)\times\text{SU}(2)$.

Now consider the operators transforming under the $\text{SU}(4)\times \text{SU}(2) \times\text{SU}(2)$ symmetry in the $(\boldsymbol{6},\boldsymbol{2},\boldsymbol{2})$ representation. After switching on the continuous Wilson line $W_6$, only 12 of them will survive:
\begin{gather}
E_{29} +E_{35}, \quad E_{30} +E_{36}, \quad E_{31} +E_{33}, \quad E_{32} +E_{34}, \quad E_{57},\notag\\
E_{58}, \quad E_{59}, \quad E_{60}, \quad E_{66}, \quad E_{67}, \quad E_{68}, \quad E_{69}
\end{gather}

Calculating the adjoint action of the Cartan generators and diagonalising the $12\times12$ matrices, we find the weight vectors:
\begin{gather}
(0,\,1/3,\,1), \quad (0,\,\text{-}1/3 \,\,1), \quad (0,\,1/3,\,\text{-}1), \quad (0,\,\text{-}1/3,\,\text{-}1),\notag\\
(1,\,1/6,\,\text{-}1), \quad (\text{-}1,\,\text{-}1/6,\,1), \quad (1,\,1/6,\,1), \quad (\text{-}1,\,\text{-}1/6,\,\text{-}1),\\
(1,\,\text{-}1/6,\,\text{-}1), \quad (1,\,\text{-}1/6,\,1), \quad (\text{-}1,\,1/6,\,\text{-}1), \quad (\text{-}1,\,1/6,\,1)\notag
\end{gather}
Using the metric in root space, we calculate the Dynkin labels by taking scalar products of the weights with the simple roots. We find the representations
\begin{equation}
(1|0 1) \quad \text{and} \quad (1|1 0)
\end{equation}
corresponding to the representations $(\boldsymbol{\overline{3}}, \boldsymbol{2})$ and $(\boldsymbol{3}, \boldsymbol{2})$ of $\text{SU}(3)\times\text{SU}(2)$.

\clearpage
\newpage

\bibliography{mybibliography}

\bibliographystyle{./utphys}

\end{document}